\documentclass[twocolumn,appendix]{aastex631}
\shorttitle{Early-type Galaxies on the MS}
\shortauthors{Koyama et al.}
\graphicspath{{./}{figures/}}

\begin{document}

\title{Early-type Galaxies on the Star Formation Main Sequence: Internal Star Formation Geometry Revealed with MaNGA and Their Environmental Origin}

\email{shuhei.koyama@nao.ac.jp}

\author{Shuhei Koyama}
\affiliation{National Astronomical Observatory of Japan, 2-21-1 Osawa, Mitaka, Tokyo 181-8588, Japan}
\author{Yusei Koyama}
\affiliation{National Astronomical Observatory of Japan, 2-21-1 Osawa, Mitaka, Tokyo 181-8588, Japan}
\affiliation{Department of Astronomical Science, Graduate University for Advanced Studies (SOKENDAI), 2-21-1 Osawa, Mitaka, Tokyo 181-8588, Japan}
\author{Takuji Yamashita}
\affiliation{National Astronomical Observatory of Japan, 2-21-1 Osawa, Mitaka, Tokyo 181-8588, Japan}
\author{Masahiro Konishi}
\affiliation{Institute of Astronomy, Graduate School of Science, The University of Tokyo, 2-21-1 Osawa, Mitaka, Tokyo 181-0015, Japan}
\author{Kosuke Kushibiki}
\affiliation{National Astronomical Observatory of Japan, 2-21-1 Osawa, Mitaka, Tokyo 181-8588, Japan}
\author{Kentaro Motohara}
\affiliation{Institute of Astronomy, Graduate School of Science, The University of Tokyo, 2-21-1 Osawa, Mitaka, Tokyo 181-0015, Japan}
\affiliation{National Astronomical Observatory of Japan, 2-21-1 Osawa, Mitaka, Tokyo 181-8588, Japan}

\begin{abstract}
Star-forming galaxies on the main sequence (MS) are often regarded as a uniform population characterized by similar global star formation properties. However, there exists a diversity in galaxy morphologies at fixed stellar mass and SFR. In this study, using spatially-resolved properties from the MaNGA final data release, we classify MS galaxies into late-type (MS-late) and early-type (MS-early). In addition, we further divide the MS-early galaxies into two distinct subgroups based on their internal star formation and stellar mass distributions within the galaxies. The first group---``MS-early\_SF''---shows centrally concentrated star formation without prominent stellar bulges and resides preferentially in dense environments, suggesting environmentally-driven evolution. The second group---``MS-early\_stellar''---exhibits significant stellar bulges with suppressed central star formation, maintains disk-like star formation patterns, and inhabits environments similar to those of late-type galaxies, indicating evolution through internal secular processes.  
Our findings demonstrate that spatially-resolved observations play critical roles in revealing the diverse evolutionary pathways hidden within galaxies that share similar global properties.

\end{abstract}


\section{Introduction} \label{sec:intro}

Galaxy evolution is influenced by a combination of many processes, including star formation, galaxy morphology, active galactic nuclei (AGN), environmental effects, and more. Understanding how these processes are related is essential for revealing the mechanisms behind galaxy evolution. Over the last few decades, large-scale surveys have provided extensive data, allowing for detailed studies of galaxy properties and their relations.
One of the most well-established relationships in galaxy evolution is the star formation main sequence (MS), a tight correlation between the stellar mass (M$_*$) and star formation rate (SFR) of star-forming galaxies across various environments \citep{Daddi07, Elbaz07, Peng10, Koyama13}. Galaxies on the MS are considered to be undergoing steady-state star formation, while those above and below the relation are classified as starbursts and passive galaxies, respectively.
Morphology is another key aspect of galaxy evolution, and it is closely related to star formation activity. The majority of star-forming galaxies on the MS have late-type morphologies, suggesting that typical star-forming galaxies are characterized by disk structures and steady, mass-dependent star formation activity. In contrast, passive galaxies are associated with early-type morphologies.
This observed connection between star formation activity and morphology suggests an evolutionary sequence where star-forming galaxies located on the MS eventually evolve into passive galaxies through the quenching of their star formation and accompanying morphological transformation \citep{Wuyts11, Bell12}.

Large integral field spectroscopy (IFS) surveys such as MaNGA \citep{Bundy15} and SAMI \citep{Croom12} provide spatially-resolved spectroscopic data for thousands of nearby galaxies, enabling statistical studies of internal star formation properties. These surveys have revealed important insights by examining galaxies based on their position relative to the MS or their morphological types.
For example, studies have found that star-forming galaxies on the MS tend to have relatively flat sSFR profiles, indicating steady-state star formation throughout their disks. In addition, central suppression of sSFR has been observed in high-mass star-forming galaxies, interpreted as a signature of inside-out quenching  \citep[e.g.,][]{Belfiore18, Ellison18}.

While previous studies have provided important insights by examining galaxies based on their MS position or morphological types, there are evidence that star-forming galaxies on the MS are not a uniform population. Galaxy morphology catalogs based on large-scale optical imaging surveys, such as Galaxy Zoo 1 and 2 \citep{Lintott11, Willett13}, have shown a variety in galaxy morphologies among MS galaxies, including a non-negligible fraction of early-type galaxies in addition to the dominant late-type population. This morphological variety is one clear example of the diversity within the MS population.
\citet{Medling18} studied resolved star formation properties of galaxies with different morphologies and reported that some early-type galaxies are found in the MS. 
While those early-type galaxies on the MS show average sSFR profiles similar to those of late-type galaxies, they also reported significant scatter in resolved properties among individual galaxies.
This scatter suggests that global properties like MS position or morphology may not fully represent the diversity of MS galaxies, and that examining internal galaxy structure and star formation distributions can reveal different populations within the MS.

Additional processes such as AGN feedback and environmental effects are particularly important to consider, as they can significantly influence both galaxy structure and star formation activity. AGN can affect the distribution and properties of star-forming gas through feedback processes \citep{Fabian12, Harrison17}, while environmental mechanisms like galaxy interactions and ram pressure stripping can reshape both the morphology and star formation activity of galaxies \citep{Boselli06, Blumenthal18}. Understanding how these processes relate to the observed diversity in star formation distributions may provide crucial insights into the different evolutionary pathways of MS galaxies.

In this paper, we investigate the diversity of MS galaxies using the final data release of the MaNGA survey. After classifying MS galaxies based on their morphology, we first quantify their spatially-resolved star formation and stellar mass distributions through S\'{e}rsic profile fitting. Based on these structural properties, we then analyze the radial profiles of specific star formation rates and resolved main sequence relations to characterize their star formation patterns in detail. Finally, we examine how these characteristics correlate with AGN activity and environment to understand the origin of the observed diversity. Through this analysis, we aim to reveal evolutionary pathways hidden within the MS diversity.

This paper is organized as follows. Section \ref{sec:data} describes the MaNGA survey data, our sample selection criteria, and the derived properties used in our analysis. Section \ref{sec:results} presents our main results, including the S\'{e}rsic profile fitting of $\Sigma_\mathrm{SFR}$ and $\Sigma_*$ maps, the comparison of mean sSFR radial profiles, and the resolved main sequence relations. The discussion is presented in Section \ref{sec:discussion}, where we explore how AGN and environment relate to the observed diversity of star formation distributions, and discuss possible evolutionary pathways. Finally, we summarize our findings in Section \ref{sec:summary}.
Throughout this paper, we adopt a $\Lambda$CDM cosmology with $H_0 = 67.7$ km s$^{-1}$ Mpc$^{-1}$ and $\Omega_M = 0.310$ \citep[Planck 2018:][]{Planck20}, and the \citet{Kroupa01} initial mass function (IMF).

\section{Data} \label{sec:data}

\subsection{The MaNGA Survey} 
\label{subsec:MaNGA}

MaNGA \citep{Bundy15} is one of the three core programs of the 4th generation of the Sloan Digital Sky Survey \citep[SDSS-IV;][]{Blanton17}. 
The survey uses the 2.5m Sloan Foundation Telescope at the Apache Point Observatory \citep{Gunn06} and employs 17 different sized integral field unit (IFU) fiber bundles, ranging from 19 to 127 fibers with a diameter of 2 arcsec. 
The spectra cover a wavelength range of 3600--10300 \AA\ at a spectral resolution of $R\sim2000$. 
The target galaxies are selected from the NASA-Sloan Atlas catalog with a redshift range of $0.01 < z < 0.15$, and is designed to be representative of the overall galaxy population at $z\sim0.03$ and covers a wide range of galaxy properties, including stellar mass, color, and morphology.
The MaNGA observing strategy employs dithered observations to ensure uniform spatial sampling across each galaxy. The final data cubes have a spatial resolution of $\sim2.5\arcsec$ (FWHM),  which corresponds to a typical physical resolution of 1.5 kpc at the median redshift of the survey ($z \sim 0.03$).

\subsection{Sample Selection}
\label{subsec:sample_selection}

\begin{figure}
\plotone{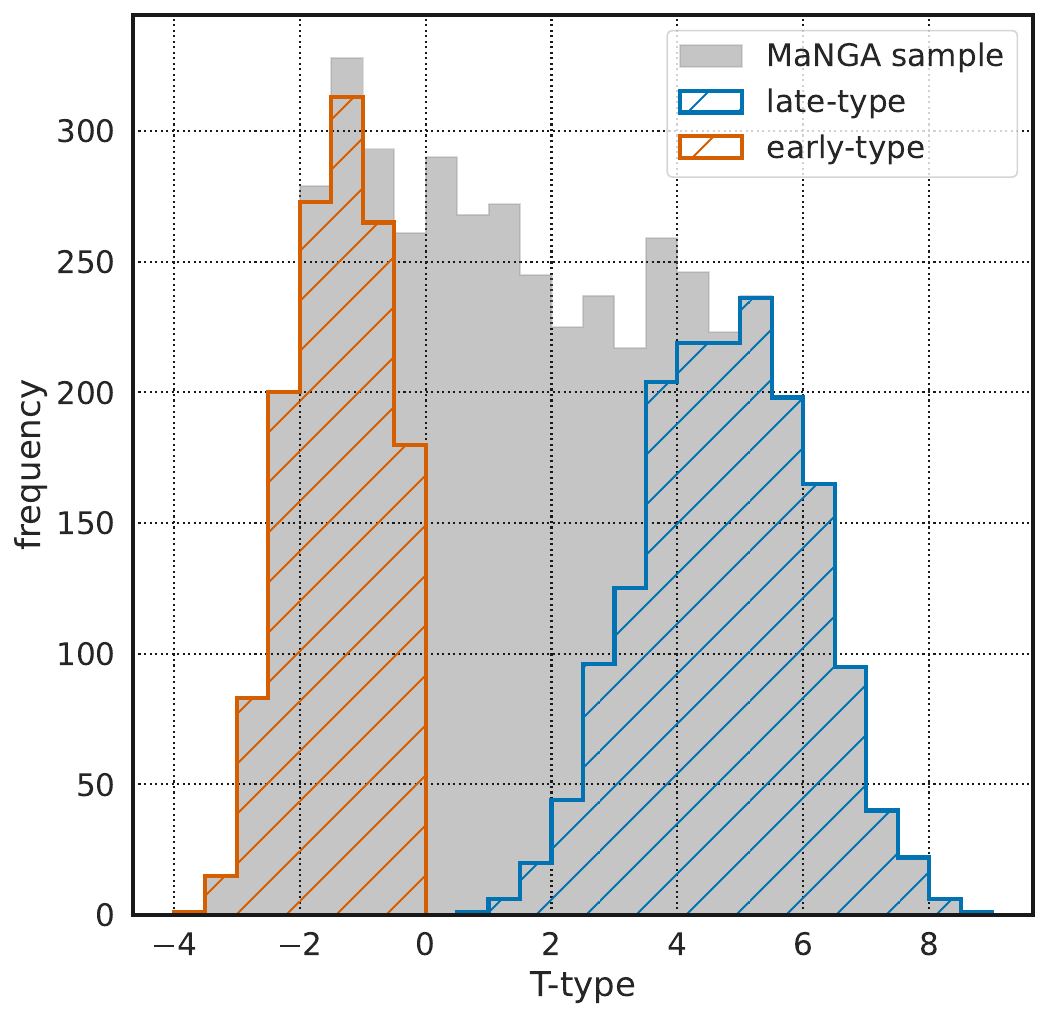}
\caption{The histogram of T-type from the MaNGA Morphology Deep Learning DR17 catalog for the full sample (gray) and the samples classified as late-type (blue) and early-type (red) galaxies based on the T-type and P\_LTG values.}
\label{TType_hist}
\end{figure}

The sample is drawn from the final data release of the MaNGA survey, the Data Release 17 \citep[DR17;][]{Abdurro'uf22}, and the Marvin software \citep{Cherinka19} is employed to access and analyze the data. 
The MaNGA data is processed through two main pipelines: the Data Reduction Pipeline \citep[DRP;][]{Law16} and the Data Analysis Pipeline \cite[DAP;][]{Westfall19}. The DRP provides sky-subtracted and flux-calibrated 3D spectra for each galaxy. The DAP provides two-dimensional maps of measured properties obtained through stellar continuum fitting and emission line measurements on the spectra produced by the DRP. In this study, we use the VOR10-MILESHC-MASTARS maps included in the DAP, in which a spatial binning is performed by the Voronoi binning algorithm \citep{Cappellari03} to achieve a signal-to-noise ratio (S/N) $\sim10$ in the g-band per spectral pixel for each bin \citep{Aguado19}.

The sample selection is based on the DAPall catalog, which is a summary table of the DAP. The selection criteria require that the maps are generated correctly as indicated by the DAPDONE flag being True, and that the data is free from quality issues (DAPQUAL$=$0). The sample is limited to galaxies with $z<0.05$, corresponding to a spatial resolution of 2.5 kpc at the typical seeing of 2.5\arcsec. Additionally, we apply NSA\_ELLPETRO\_BA $>0.5$ to select galaxies with an elliptical petrosian B/A ratio greater than 0.5. The sample consists of 4706 galaxies.

We perform morphological classification using the MaNGA Morphology Deep Learning DR17 catalog \citep{Dominguez22}, which provides deep learning-based morphological classifications trained on two visually-based morphological catalogs \citep{Nair10, Willett13}. The T-type is a numerical classification scheme that ranges from -4 to 9, with positive values indicating spiral galaxies of increasingly later types. The P\_LTG parameter represents the probability of a galaxy being late-type, with values ranging from 0 (early-type) to 1 (late-type). 
In this paper, we combine these parameters with conservative thresholds, to select galaxies with distinct morphologies (i.e. early-type and late-type). We define "early-type" galaxies as those with T-type $\leq$ 0 and P\_LTG $< 0.1$, and "late-type" galaxies as those with T-type $> 0$ and P\_LTG $> 0.9$.
This classification yields 1697 late-type and 1330 early-type galaxies. 
Since visually-based morphological classifications can be affected by spatial resolution and color variations \citep[e.g.,][]{Masters21,Ren24,Salvador24,Foster25}, we verify the reliability of our morphological classification by examining the correlation between T-type and S\'{e}rsic index from NSA r-band measurements, finding a strong anti-correlation (correlation coefficient $r = -0.7$) as expected. The S\'{e}rsic index is more robust against poor spatial resolution and agnostic to color. We calculate residuals from the T-type vs S\'{e}rsic index relation and confirm that these residuals show only weak correlations with redshift, SDSS g-r color, or spatial resolution (FWHM/Re) (correlation coefficients $r = 0.05, 0.007$, and -0.2, respectively), indicating that systematic effects from distance, color, or spatial resolution have minimal impact on our morphological classifications within our sample. In Figure~\ref{TType_hist}, we show the histogram of the T-type for the full sample (in gray histogram) and the samples classified as late-type and early-type galaxies (in red and blue shaded histograms, respectively).

\subsection{Derived properties}
\label{subsec:derived_prop}

\begin{figure*}
\plotone{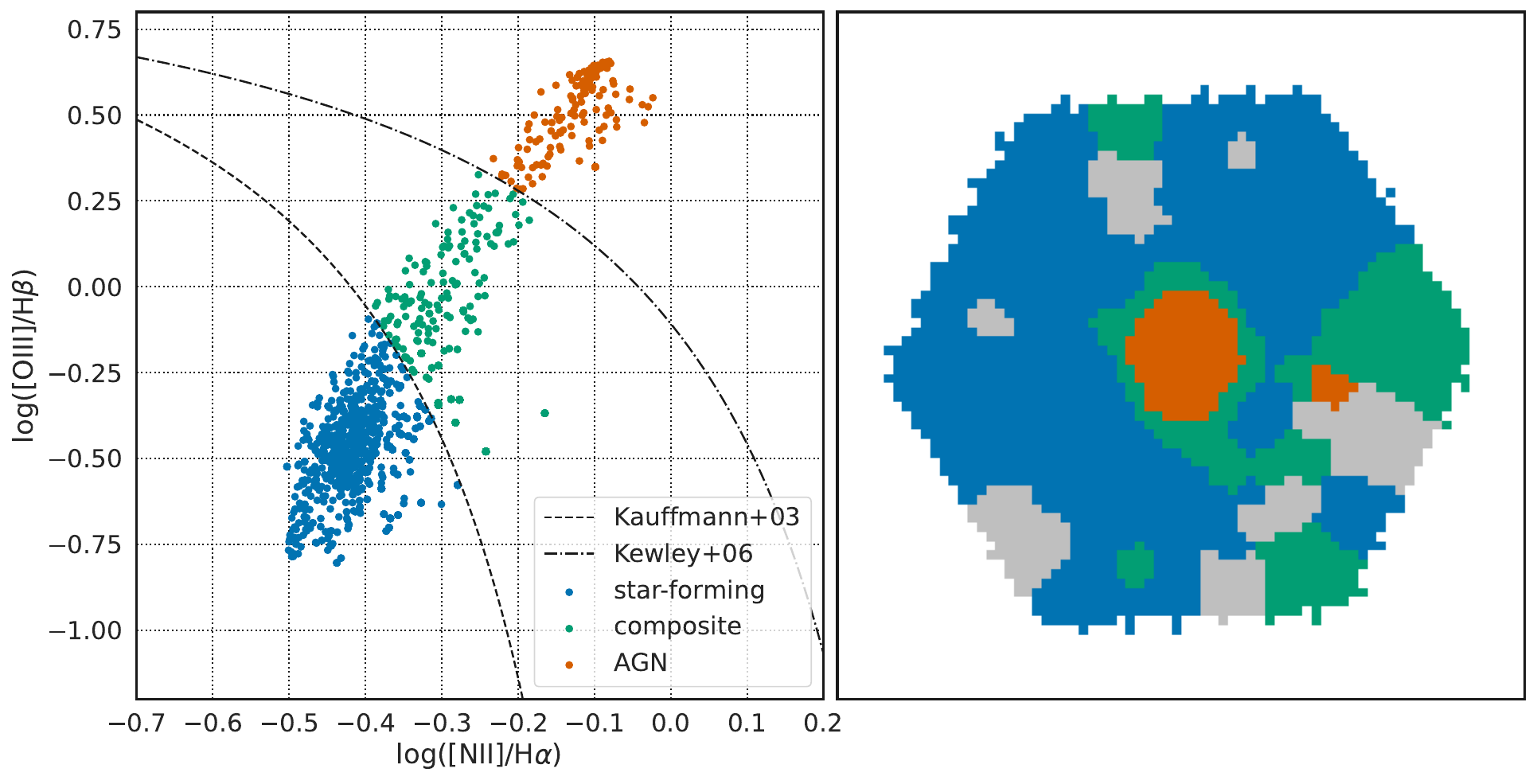}
\caption{An example of the resolved BPT diagram. The left panel shows the [\ion{N}{2}]/H$\alpha$ vs. [\ion{O}{3}]/H$\beta$ line ratios for individual spaxels, with the demarcation lines from \citet{Kauffmann03} (dashed line) and \citet{Kewley06} (dash-dotted line). Spaxels are classified as star-forming (blue), composite (green), or AGN (red) based on their locations relative to these lines. The right panel displays the spatial distribution of the classified spaxels within the galaxy using the same color scheme. Spaxels with S/N $< 3$ in H$\alpha$, H$\beta$, [\ion{N}{2}], or [\ion{O}{3}] are shown in grey.}
\label{BPT}
\end{figure*}

The emission line measurements are taken from the emline\_gflux maps in the MaNGA DAP. These maps are derived through a two-step process in the DAP pipeline \citep{Westfall19}. First, the stellar continuum is modeled and subtracted from each spectrum using a combination of stellar templates from the MILES stellar library \citep{SanchezBlazquez06}. This step accounts for stellar absorption features that could affect emission line measurements, particularly for Balmer lines. After the continuum subtraction, the emission lines are fit using Gaussian profiles.
Using these emission line measurements, we derive the star formation rate surface density ($\Sigma_{\mathrm{SFR}}$) from the H$\alpha$ and H$\beta$ emission line maps. First, we calculate the Balmer decrement (H$\alpha$/H$\beta$ ratio) to estimate dust attenuation at each spaxel, assuming Case B recombination and the \citet{Cardelli89} attenuation curve. The dust-corrected H$\alpha$ luminosity is then converted to $\Sigma_\mathrm{SFR}$ using the calibration from \citet{Kennicut94}.

We also construct the resolved BPT \citep{Baldwin81} diagram to distinguish star-forming regions from other ionization sources on a spaxel-by-spaxel basis using H$\alpha$, H$\beta$, [\ion{N}{2}], and [\ion{O}{3}] maps. In Figure \ref{BPT}, we show an example of the resolved BPT diagram. Based on \citet{Kauffmann03} and \citet{Kewley06}, we distinguish star-forming, composite and AGN spaxels. Throughout this paper, we exclude spaxels classified as AGN or composite from our analysis of star formation properties ($\Sigma_\mathrm{SFR}$ and SFR) to measure pure star formation activity uncontaminated by other ionization sources.

The stellar mass surface density ($\Sigma_*$) is obtained from the MaNGA Firefly Stellar Populations value-added catalog \citep{Goddard17,Neumann22}. This catalog provides spatially resolved stellar population properties derived using the FIREFLY spectral fitting code \citep{Wilkinson17}, which performs a full spectral fitting to the stellar continuum using the MaStar stellar population models \citep{Maraston20}.

Finally, we calculate the total star formation rate (SFR) by summing up the $\Sigma_\mathrm{SFR}$ of the spaxels classified as star-forming for each galaxy. The total stellar mass (M$_*$) is derived by summing up the $\Sigma_*$ from all spaxels. To validate our measurement methods, we compare our measurements with those from the GALEX-SDSS-WISE Legacy Catalog 2 \citep[GSWLC-2;][]{Salim16,Salim18}, which derives M$_*$ and SFR through UV/optical/mid-IR SED fitting. 
The measurements show tight correlation with scatter of 0.2 dex for both SFR and M$_*$. We find that our SFR measurements are typically 0.1 dex higher and our M$_*$ measurements are 0.2 dex lower than the GSWLC-2 values. 
This offset in SFR may be related to several methodological factors, including the different timescales measured by H$\alpha$ versus UV+IR techniques \citep{Kennicutt12}, spatially-resolved dust corrections, and differences in calibration assumptions. While we cannot determine the contribution of each factor, the overall offset is relatively small compared to the scatter between the two measurements (0.2 dex).
For the stellar mass, this offset is consistent with \citet{Neumann22}, who reported that Firefly-derived total stellar masses (MaStar model) are typically 0.05-0.32 dex lower than other catalog masses due to differences in stellar population models and fitting methodologies. Most importantly, since our sample selection is based on the relative position of galaxies in the M$_*$-SFR parameter space using our own measurements, these systematic offsets do not affect our analysis.

Furthermore, to assess potential biases in our SFR measurements from excluding AGN/composite spaxels, we perform an additional test. For MS galaxies hosting AGN/composite spaxels, we compare two measurements of total SFR. The first includes only star-forming spaxels, while the second additionally includes AGN and composite spaxels. This second measurement provides an upper limit on the potential star formation activity that might be obscured by AGN emission. The difference between these measurements has a median of 0.14 dex, suggesting that the exclusion of AGN/composite spaxels does not significantly affect our measurements of SFR.

\subsection{MS-late and MS-early classification}
\label{subsec:MS_classification}

\begin{figure}
\plotone{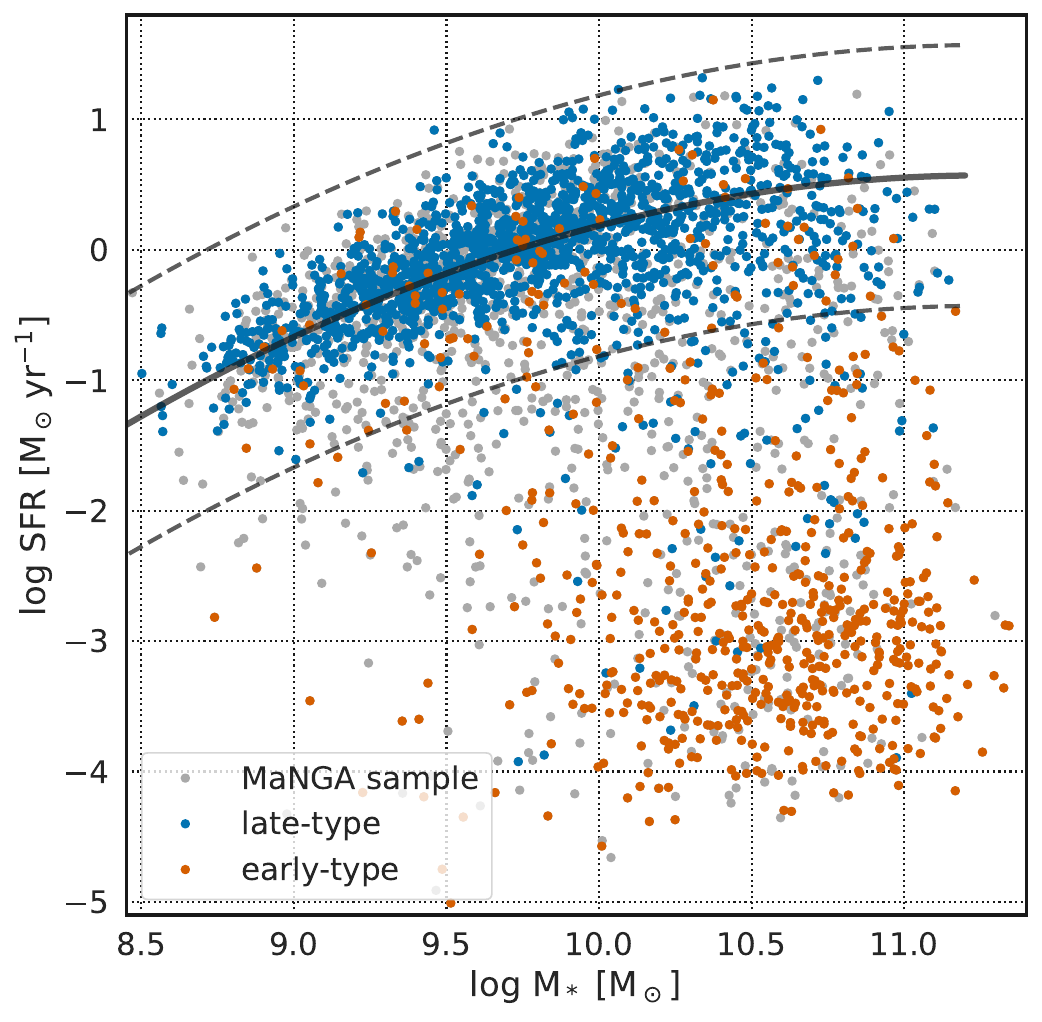}
\caption{The relation between total stellar mass (M$_\star$) and total star formation rate (SFR) for the morphology classified sample. Blue circles show late-type galaxies, while red circles show early-type galaxies. The black solid line indicates the best-fit star-formation main sequence relation. The dashed lines show the $\pm1$ dex scatter around the main sequence relation.}
\label{M_SFR}
\end{figure}

\begin{figure*}
\plotone{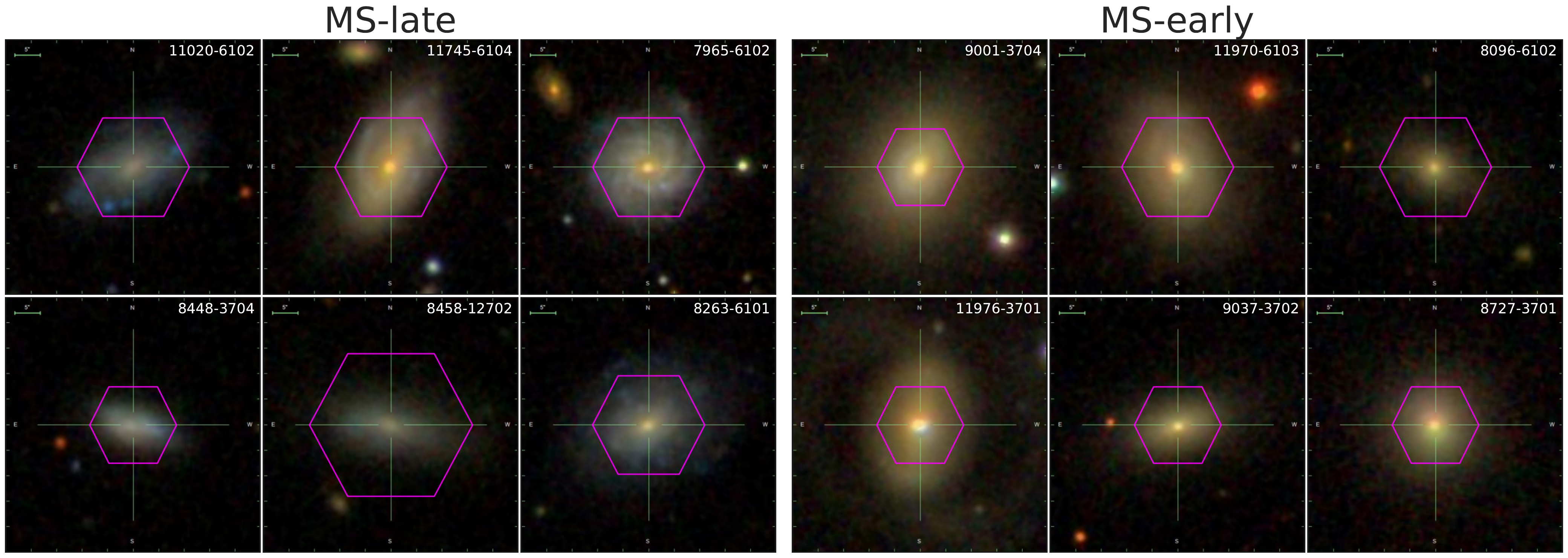}
\caption{SDSS $gri$ color composite images for six randomly selected MS-late galaxies (left panels) and MS-early galaxies (right panels). Each image is 50\arcsec $\times$ 50\arcsec\ in size, with the MaNGA footprint overlaid as a pink outline.}
\label{SDSSimages}
\end{figure*}

\begin{figure}
\plotone{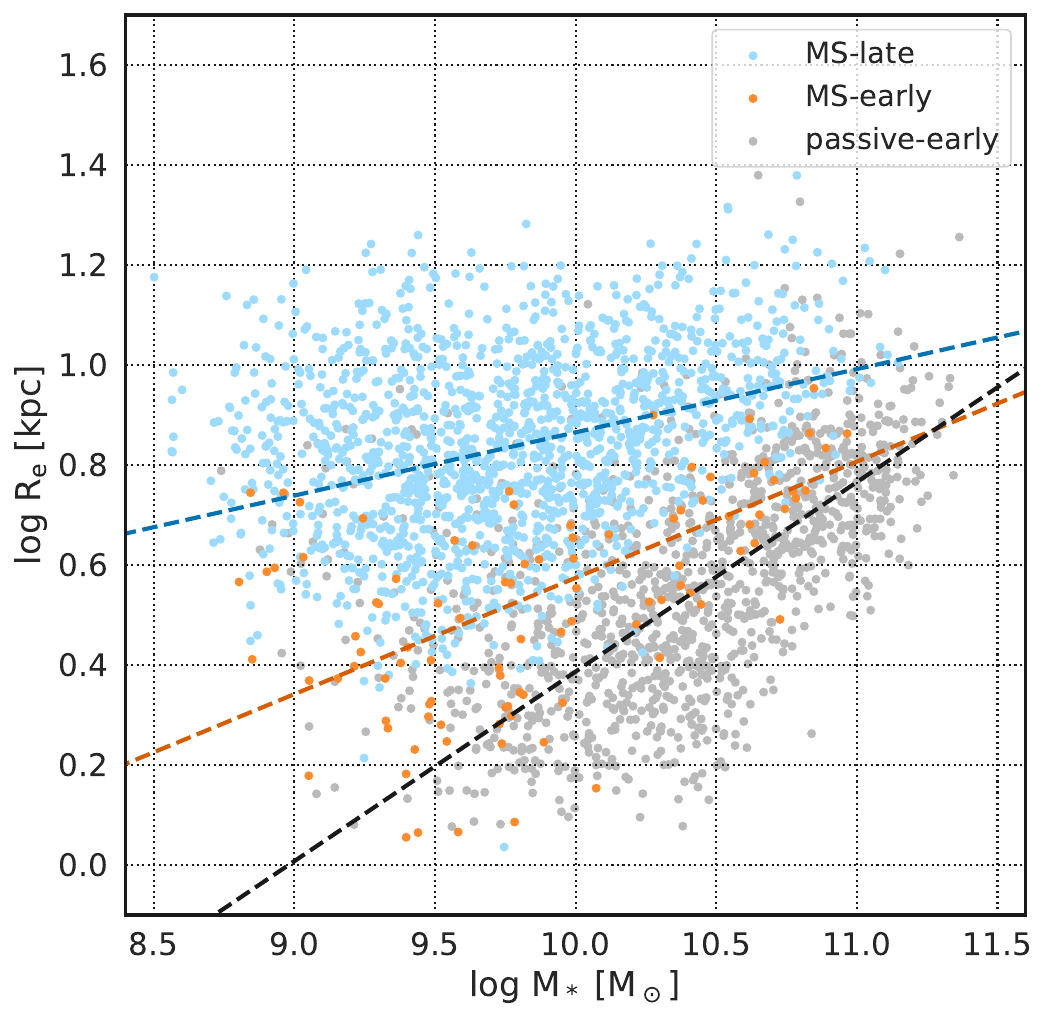}
\caption{ 
The relation between stellar mass and effective radius for the MS-late (blue circles), MS-early (red circles), and passive early-type (grey circles) galaxies. Lines show the best-fit linear relations for passive early-type (black dashed), MS-early (red dashed), and MS-late (blue dashed) galaxies.
}
\label{M_Re}
\end{figure}

In Figure \ref{M_SFR}, we show the relation between M$_*$ and SFR for late-type and early-type galaxies.
The star-formation main sequence (MS) is defined by fitting a second-order polynomial to galaxies with EW(H$\alpha$) $>$ 5. The solid line represents the best-fitting line, while the dashed lines indicate the $\pm$1 dex ($\sim 3\sigma$) range. As expected, most late-type galaxies are located along the MS, while the majority of early-type galaxies are passive. However, it should be noted that a small, but non-negligible fraction of early-type galaxies are located on the MS.
We classify galaxies falling within the dashed lines in Figure \ref{M_SFR} as MS galaxies, and refer to those with late-type and early-type morphologies as the MS-late (1583 galaxies) and the MS-early (97 galaxies), respectively.
Figure \ref{SDSSimages} presents examples of SDSS images of MS-late and MS-early galaxies. MS-late galaxies show typical disk-like morphology, while MS-early galaxies appear to have smooth, elliptical-like structures despite their classification as MS galaxies.

We also note that the distribution of effective radius is clearly different between MS-late and MS-early galaxies. Figure~\ref{M_Re} shows the relation between stellar mass and effective radius for our sample, where passive early-type galaxies are defined as early-type galaxies with SFRs below the MS as shown in Figure~\ref{M_SFR}. We find that MS-late and passive early-type galaxies in our sample show a trend similar to the well-known mass-size relation \citep{vanderWel14}, where MS-late galaxies have systematically larger sizes than passive early-type galaxies at fixed stellar mass.
Interestingly, MS-early galaxies show an intermediate mass-size relation between MS-late and passive early-type galaxies, though slightly closer to the passive early-type distribution.

In the following analyses, we investigate the diversity within the MS population by comparing the MS-late and MS-early galaxies as a starting point. The main aim of this study is to study the diversity of \textit{internal} galaxy properties among the MS galaxies, through the spatially-resolved analyses.

\section{Results} \label{sec:results}

\subsection{S\'{e}rsic Profile Analysis of $\Sigma_\mathrm{SFR}$ and $\Sigma_*$ Maps} \label{subsec:sersic_analysis}

Our analysis aims to characterize and compare the spatial distributions of star formation activity in MS-late and MS-early galaxies. To achieve this, as a first step, we quantify the radial profiles of star formation rate surface density ($\Sigma_\mathrm{SFR}$) and stellar mass surface density ($\Sigma_*$) for individual galaxies using the S\'{e}rsic profile fitting.
The S\'{e}rsic profile is defined as:
\begin{equation}
    I(r) = I_e \exp\left\{-b_n\left[\left(\frac{r}{R_e}\right)^{1/n} - 1\right]\right\},
\end{equation}
where $I(r)$ is the intensity at radius $r$, $I_e$ is the intensity at the effective radius $R_e$, $n$ is the S\'{e}rsic index, and $b_n$ is a function of $n$ that ensures half of the total luminosity is enclosed within $R_e$. 
Lower values of $n$ indicate more uniform or late-type like distributions, while higher values represent more centrally concentrated profiles typical of bulges or elliptical galaxies.
This approach allows us to simplify the complex spatial information contained in the IFS data into a set of parameters that can be easily compared across our sample, and allows us to study how star formation activity is distributed within galaxies across different morphologies.

\begin{figure*}
\plotone{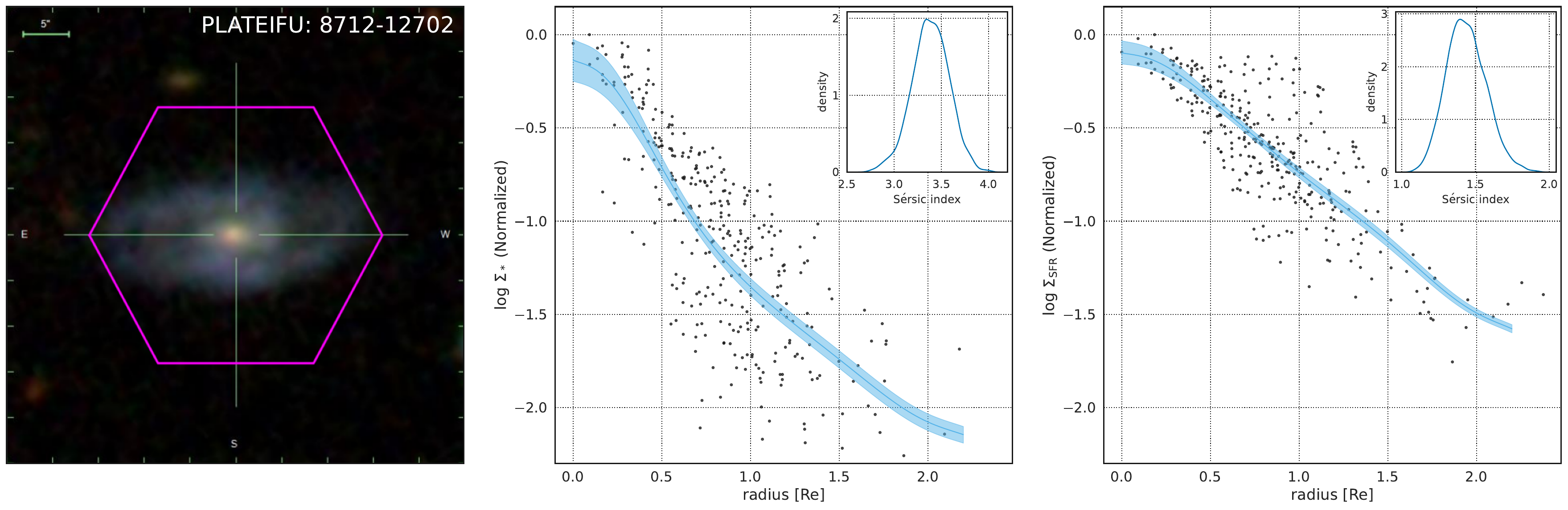}
\caption{
Examples of S\'{e}rsic profile fitting for stellar mass surface density ($\Sigma_*$, middle) and star formation rate surface density ($\Sigma_\mathrm{SFR}$, right) maps of the same galaxy. The left panel shows the SDSS image of the galaxy. Black circles in the middle and right panels represent the observed radial profiles normalized by their maximum values. Cyan lines show the best-fitting PSF-convolved S\'{e}rsic models with shaded regions indicating the 95\% confidence intervals. The insets show the probability density distributions of the S\'{e}rsic indices derived from the fitting.
}
\label{fitting_example}
\end{figure*}

To perform the S\'{e}rsic profile fitting, we proceed with the following steps. First, we use the elliptical coordinates of each spaxel from the galaxy center (spx\_ellcoo\_r\_re). These coordinates account for the ellipticity and position angle of each galaxy.
To account for the point spread function (PSF) effects, we convolve the S\'{e}rsic profile models with the MaNGA PSF. The PSF is well approximated by a Gaussian with a FWHM of 2.54\arcsec\ \citep{Law16}. To match the spatial binning of the observed maps, we apply the Voronoi binning scheme used in the MaNGA DAP to these PSF-convolved S\'{e}rsic maps using the BINID map.
We then fit these binned, PSF-convolved S\'{e}rsic models to the observed $\Sigma_\mathrm{SFR}$ and $\Sigma_*$ maps using a Markov Chain Monte Carlo (MCMC) approach within a Bayesian framework. This method provides probability distributions for the S\'{e}rsic indices rather than single best-fit values, allowing us to better account for uncertainties in our analysis. 
Figure \ref{fitting_example} demonstrates examples of our S\'{e}rsic profile fitting for both $\Sigma_*$ and $\Sigma_\mathrm{SFR}$ maps of the same galaxy. The radial profiles reveal different shapes with the $\Sigma_*$ distribution showing a steeper decline than the $\Sigma_\mathrm{SFR}$ distribution. This difference is also confirmed by the probability density distributions of the S\'{e}rsic indices, where $\Sigma_*$ shows a higher S\'{e}rsic index than that of $\Sigma_\mathrm{SFR}$.

\begin{figure*}
\plotone{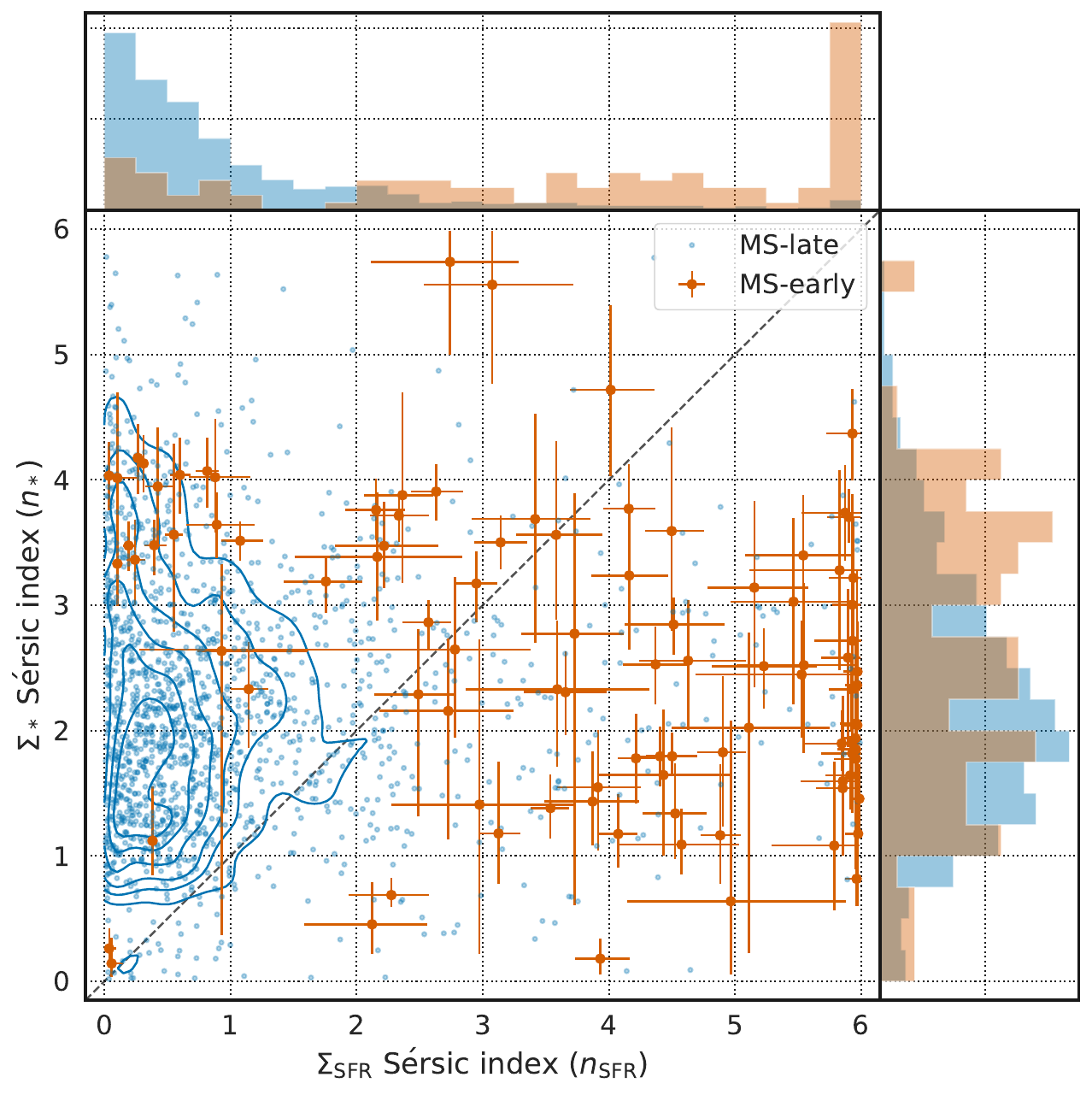}
\caption{
Distribution of S\'{e}rsic indices derived from $\Sigma_\mathrm{SFR}$ and $\Sigma_*$ maps. In the main panel, the blue contours and circles represent the distribution of MS-late galaxies, while red circles show MS-early galaxies. Error bars on the MS-early points indicate the 95\% confidence intervals from the S\'{e}rsic fitting. The dashed line represents where the S\'{e}rsic indices of $\Sigma_\mathrm{SFR}$ and $\Sigma_*$ are equal. Top and right panels show the normalized histograms of the S\'{e}rsic indices for $\Sigma_\mathrm{SFR}$ and $\Sigma_*$, respectively, following the same color scheme as the main panel.
}
\label{sersic_2d}
\end{figure*}

Figure \ref{sersic_2d} presents the joint distribution of S\'{e}rsic indices derived from the $\Sigma_\mathrm{SFR}$ (n$_\mathrm{SFR}$) and the $\Sigma_*$ maps (n$_*$). The MS-late galaxies show a consistent pattern in their $\Sigma_\mathrm{SFR}$ profiles with low S\'{e}rsic indices ($n_\mathrm{SFR} < 1$), indicating extended star formation activity across their disks. Their $\Sigma_*$ profiles exhibit a broader distribution peaked around $n_* \sim 2$, suggesting that while all the MS-late galaxies share similar extended star formation patterns, they show more variation in their stellar mass concentration. This variation likely reflects differences in the relative prominence of central bulge components.

In contrast, the MS-early galaxies show a diversity in their structural properties. While some MS-early galaxies show concentrated stellar mass distribution (high $n_*$) with extended star formation (low $n_\mathrm{SFR}$), others exhibit the opposite trend with more concentrated star formation activity despite relatively lower stellar mass concentration. 
For further analysis of these structural variations, we classify MS-early galaxies into two subgroups. Using the line where $n_*$ and $n_\mathrm{SFR}$ are equal as a reference (broken line in Figure \ref{sersic_2d}), we define the MS-early galaxies with $n_* > n_\mathrm{SFR}$ as the ``MS-early\_stellar", and those with $n_* < n_\mathrm{SFR}$ as the ``MS-early\_SF". Among the total 97 MS-early galaxies in our sample, 63 galaxies are classified as MS-early\_SF and 34 galaxies as MS-early\_stellar.

We note that some MS-early galaxies are distributed around $n_\mathrm{SFR} \sim 6$, because of the limitation of our fitting procedure and the image quality of the data. This situation happens when the central star formation is more compact than the PSF size, particularly when the PSF-convolved S\'{e}rsic profile approaches the PSF shape. 
While it is difficult to constrain the exact values of their S\'{e}rsic indices in this case, this does not affect our main conclusion. We regard these galaxies indeed harbor centrally concentrated star formation, and thus distinct populations from those with extended star formation over the galaxy disks.

We also note that for galaxies with central regions classified as AGN or composite, our S\'{e}rsic profile fitting uses only the available star-forming spaxels, creating gaps in their central $\Sigma_\mathrm{SFR}$ maps. These AGN host galaxies tend to have lower $n_\mathrm{SFR}$ values. This trend could result from potential star formation activity that may be present but obscured by AGN emission in the central regions, causing $n_\mathrm{SFR}$ to be underestimated in these galaxies. However, when we reproduce our main result's figures using only non-AGN galaxies (Appendix \ref{appendix:noAGN}), the characteristic differences between the galaxy groups remain consistent.

\subsection{Mean Specific SFR Radial Profiles}
\label{subsec:radial_profiles}

\begin{figure*}
\plotone{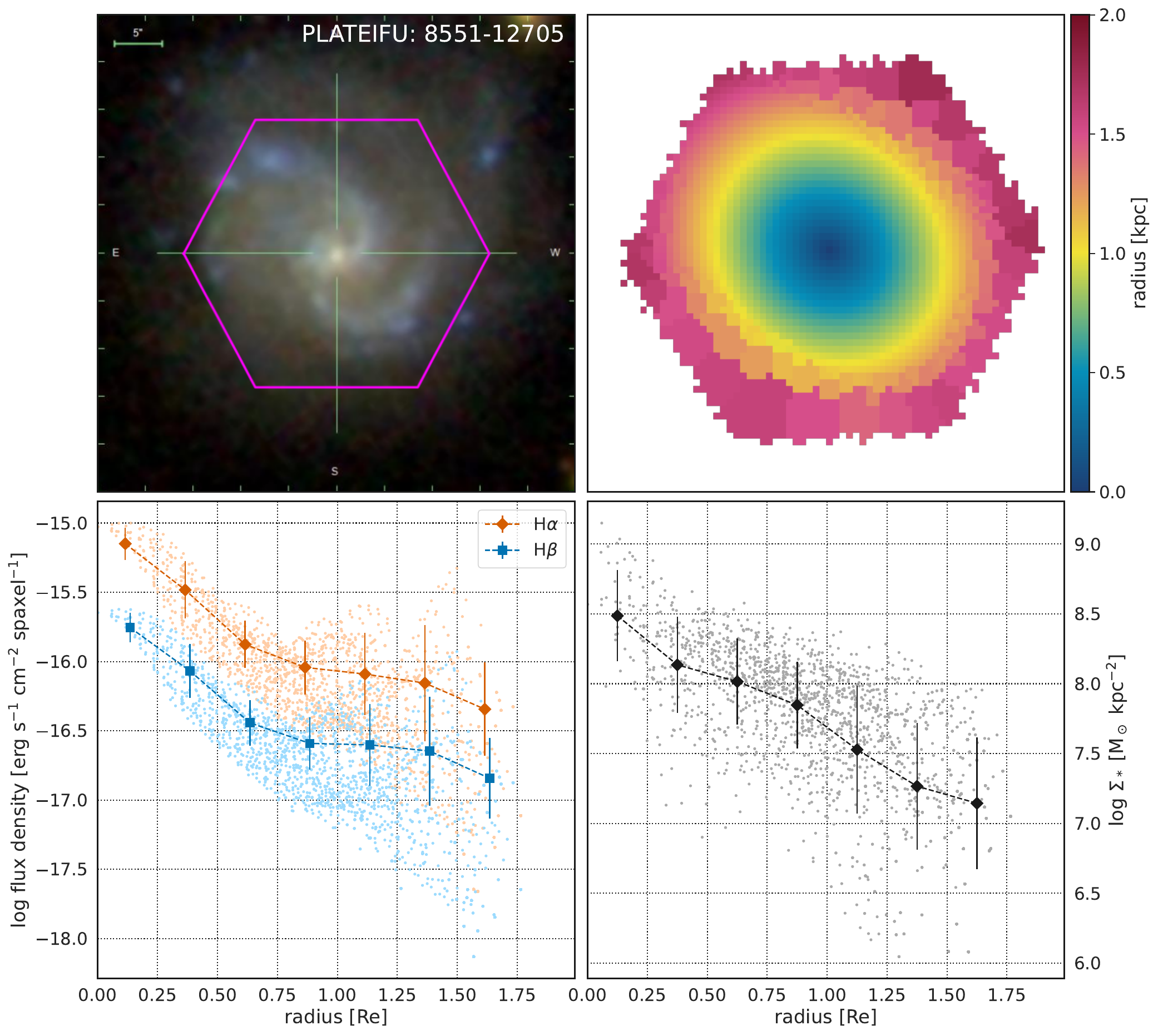}
\caption{
Example of the radial profile calculation process for a single galaxy. The top left panel shows the SDSS image of the galaxy, while the top right panel displays the corresponding MaNGA elliptical coordinate map. The bottom panels present the derived radial profiles: the H$\alpha$ and H$\beta$ profiles are shown in the bottom left (blue and red circles, respectively), and the stellar mass density profile is shown in the bottom right (black circles). In the radial profile panels, the diamonds with dashed lines and error bars indicate the weighted means and standard deviations in each radial bin.}
\label{MaNGAexample}
\end{figure*}

To investigate differences in the spatial distributions of star formation activity between MS-late and two MS-early subgroups, we compare their mean radial profiles of specific star formation rates (sSFR). For this purpose, we calculate weighted mean radial profiles for the H$\alpha$, H$\beta$, and stellar mass density maps using the elliptical coordinate maps provided in the MaNGA DAP for each galaxy. The coordinate is expressed in units of effective radius-normalized distance [R$_e$] (bin\_lwellcoo\_r\_re). We calculate the weighted mean in bins of 0.5 R$_e$ unit to obtain radial profiles for each galaxy (Figure \ref{MaNGAexample}). For the H$\alpha$ and H$\beta$ maps, we include all spaxels classified as star-forming or unclassified due to weak emission lines, while excluding those classified as composite or AGN. With this approach, we can properly include low S/N spaxels and avoid the bias toward the brightest star-forming regions. 

Next, we compute the sSFR radial profiles from the weighted mean radial profiles of H$\alpha$, H$\beta$, and stellar mass density, following the same procedure as described in Section \ref{subsec:derived_prop}. Before estimating the mean radial profiles, we normalize the sSFR radial profile of each galaxy by its mean sSFR value. This step is performed to remove the variations in the overall offset of individual galaxy profiles, allowing us to focus on the shape of the profiles. Here, we select normalization by the mean sSFR value rather than at a specific radius to minimize the impact of localized variations, thereby reducing scatter in the combined profiles.

We then divide both MS-late and MS-early into two stellar mass groups: low-mass ($8.5 \leq \log(M/M_\odot) < 9.75$) and high-mass ($9.75 \leq \log(M/M_\odot) < 11.0$), to account for potential stellar mass dependencies. Finally, we calculate the mean radial profiles for the $2\times3$ groups defined by stellar mass (low-mass or high-mass) and morphology (MS-late, MS-early\_SF and MS-early\_stellar).

\begin{figure*}
\plotone{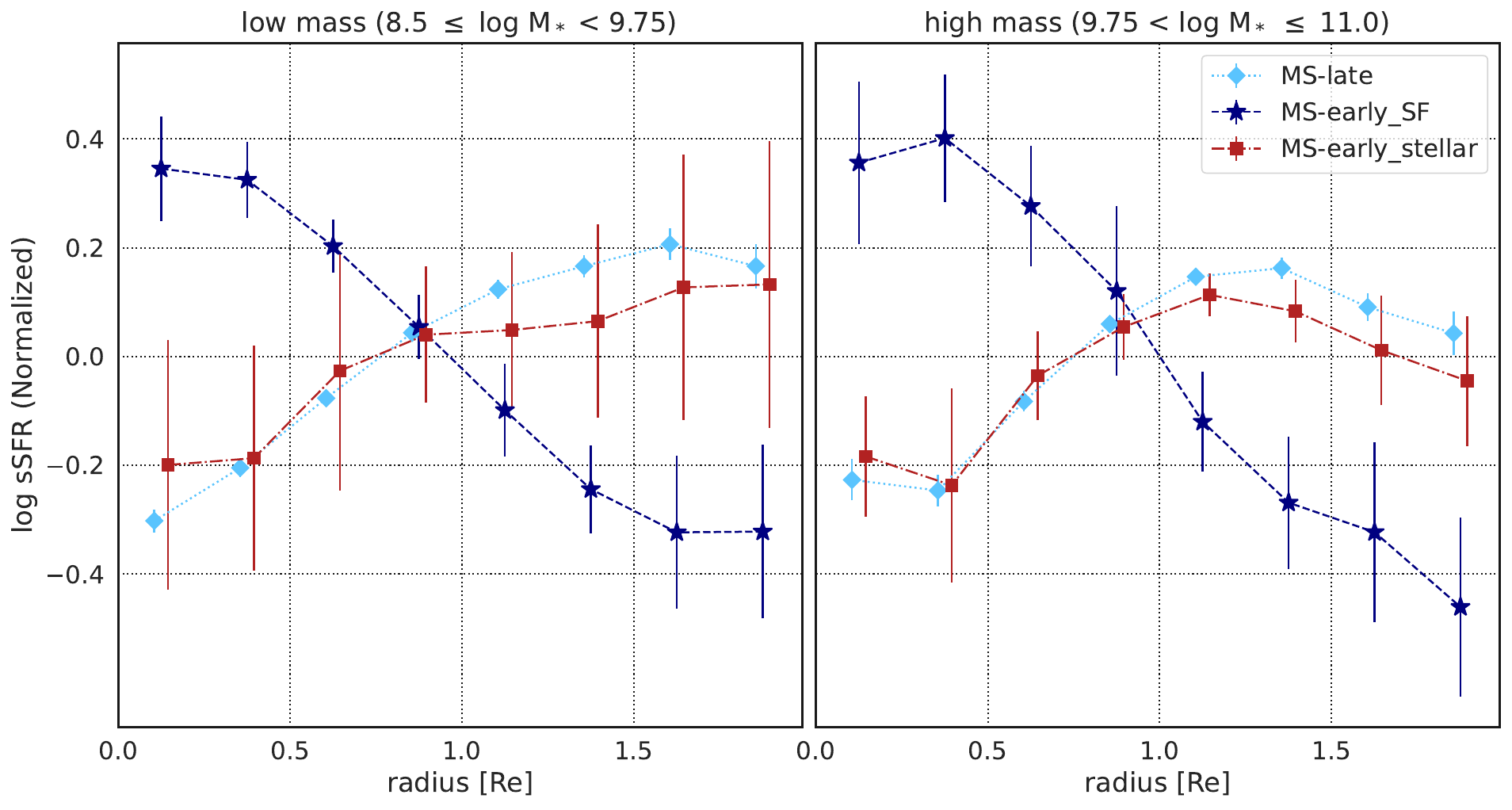}
\caption{
The mean sSFR radial profiles for the three galaxy groups: MS-late (light blue diamonds), MS-early\_SF (navy stars), and MS-early\_stellar (red squares). The error bars represent the 95\% confidence intervals of the mean, assuming a t-distribution. The mean profiles are shown separately for low-mass ($8.5 \leq \log(M/M\odot) < 9.75$; left panel) and high-mass ($9.75 \leq \log(M/M_\odot) < 11.0$; right panel) galaxies. 
}
\label{SSFRprofile}
\end{figure*}

Figure \ref{SSFRprofile} presents the mean sSFR radial profiles. The error bars represent the 95\% confidence intervals of the mean, assuming a t-distribution.
Regardless of the stellar mass range, the MS-late galaxies show a profile with lower sSFR in their central regions and higher sSFR in their outer regions. This trend is consistent with previous studies focusing on MS galaxies (or late-type galaxies). Specifically, these galaxies have active star formation in their disk, while the central region have less active star formation compared to the disk, which is often cited as evidence for inside-out quenching.
Interestingly, the MS-early\_stellar galaxies display radial profiles remarkably similar to those of MS-late galaxies. This similarity suggests that MS-early\_stellar galaxies might represent a more evolved state of MS-late galaxies, where the stellar bulge has grown more prominent while maintaining a similar pattern of star formation activity.
In contrast, MS-early\_SF galaxies show the opposite trend, with significantly enhanced sSFR in their center that rapidly decreases towards the outer region. This steep gradient is consistent with their high $n_\mathrm{SFR}$ found in Section \ref{subsec:sersic_analysis}.

\subsection{Resolved Main Sequence}
\label{subsec:resolved_ms}
\begin{figure*}
\plotone{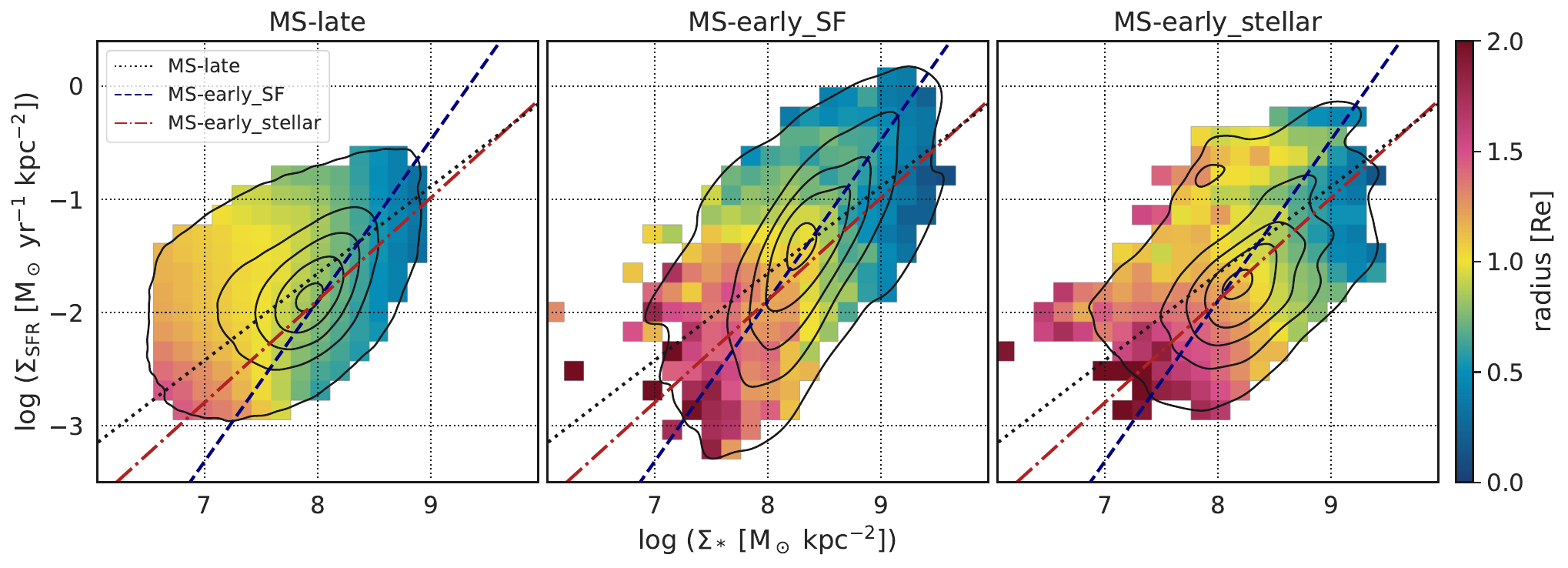}
\caption{
The resolved main sequence (the relation between log $\Sigma_*$ and log $\Sigma_\mathrm{SFR}$) for the three galaxy groups. From left to right, the panels show the distributions for MS-late, MS-early\_SF, and MS-early\_stellar galaxies. In each panel, the contours represent the 2D histogram of spaxel values, and the overlaid colormaps show the mean radius [R$_e$] calculated for each bin in the log $\Sigma_*$-log $\Sigma_\mathrm{SFR}$ space with bin sizes of $\sim0.17$ dex.
The three lines indicate the results of orthogonal fitting to the distributions. The black dotted line corresponds to the MS-late, the blue dashed line corresponds to the MS-early\_SF, and the red dot-dashed line corresponds to the MS-early\_stellar.}
\label{resolvedMS}
\end{figure*}

Finally, we compare the relation between $\Sigma_*$ and $\Sigma_\mathrm{SFR}$, known as the resolved main sequence. The $\Sigma_\mathrm{SFR}$ values used in this analysis are dust-corrected as described in Section \ref{subsec:derived_prop}. Figure \ref{resolvedMS} presents the resolved main sequence for the MS-late, MS-early\_SF, and MS-early\_stellar. To create the resolved main sequence, we plot each spaxel value on the $\Sigma_*$ and $\Sigma_\mathrm{SFR}$ plane, and then calculate 2D histogram, which is represented by the black contour lines.

As shown in Figure \ref{resolvedMS}, in the case of the MS-early\_SF, the slope of the distribution appears steep. Orthogonal fitting reveals that the slope is $1.42 \pm 0.03$, which is significantly steeper compared to the MS-late galaxies ($0.771 \pm 0.002$). This suggests that, even compared to typical star-forming galaxies, the central star formation activity in the MS-early\_SF is intense, indicating the presence of a central starburst. Conversely, the outer regions exhibit less active star formation, leading to overall SFR that is comparable to the MS.

On the other hand, the slope of the distribution for the MS-early\_stellar is $0.91 \pm 0.02$, moderately steeper than that of the MS-late galaxies but much less than MS-early\_SF. This steeper slope appears to be driven by moderately reduced $\Sigma_\mathrm{SFR}$ in outer regions rather than enhanced central star formation. These differences suggest that MS-early\_stellar galaxies maintain star formation activity largely similar to MS-late galaxies, with only modest variations in their outer regions.
As noted in Section \ref{subsec:sersic_analysis}, the $\Sigma_*$ distribution in the MS-early\_stellar is more centrally concentrated compared to both the MS-late and the MS-early\_SF.
This suggests that the MS-early\_stellar galaxies are dominated by stellar bulges, similar to ellipticals, while maintaining a spread-out distribution of star formation activity, that is a consistent result with the sSFR radial profiles in Section \ref{subsec:radial_profiles}. 

\section{Discussion}
\label{sec:discussion}
Our analysis revealed that MS-early galaxies can be further classified into two distinct subgroups: MS-early\_SF and MS-early\_stellar. These subgroups exhibit different star formation geometry despite both being located on the MS. To understand the origin of these subgroups, we further examine these populations in the perspectives of AGN activity and environmental effects.

\begin{figure}
\plotone{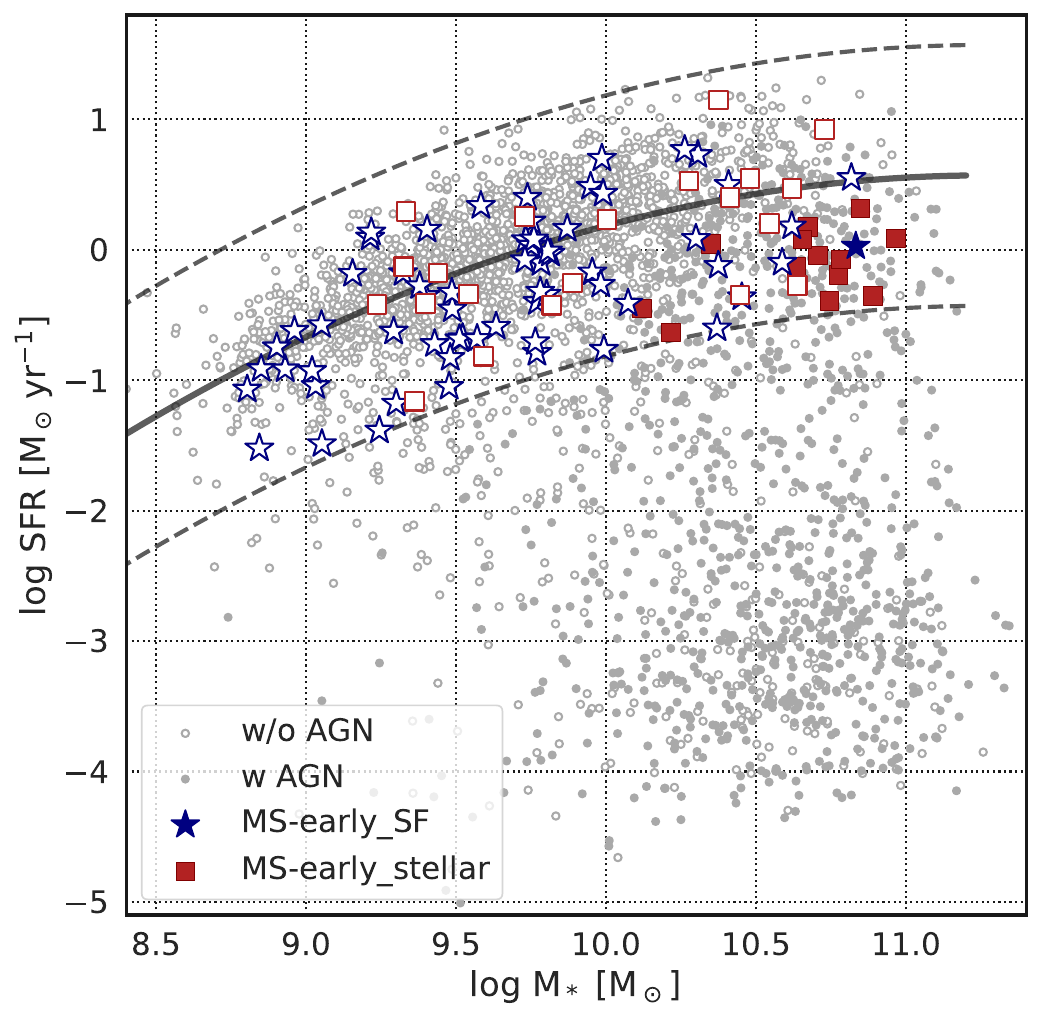}
\caption{The relation between stellar mass (M$_*$) and star formation rate (SFR). Gray circles show the full sample, while blue stars and red squares represent the MS-early\_SF and the MS-early\_stellar galaxies, respectively. Filled symbols indicate AGN host galaxies, while open symbols represent non-AGN galaxies.}
\label{M_SFR_AGN}
\end{figure}

\begin{figure*}
\plotone{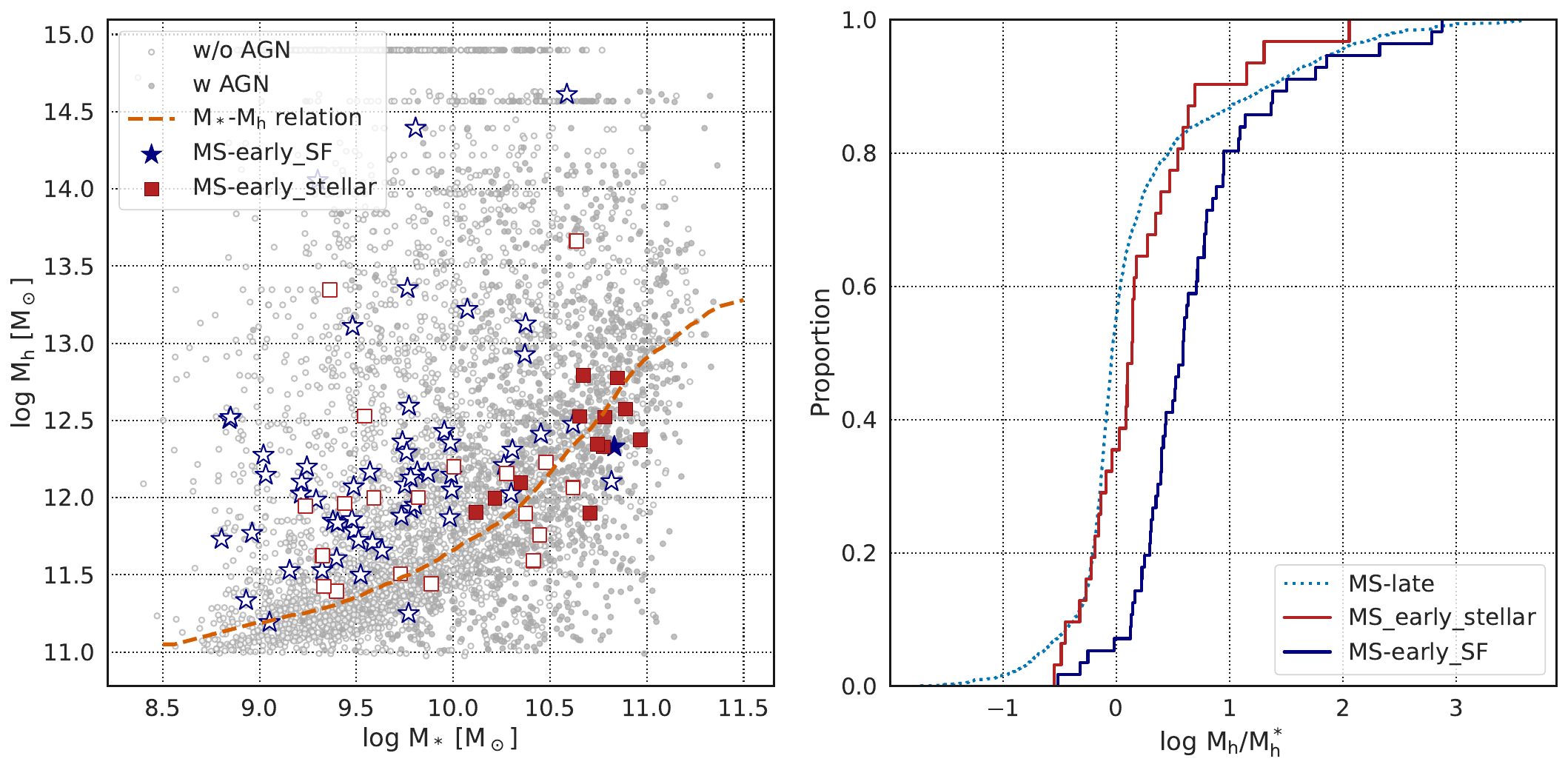}
\caption{Left: The relation between stellar mass and halo mass. Gray circles show the full sample, while blue stars and red squares represent the MS-early\_SF and the MS-early\_stellar galaxies, respectively. Filled symbols indicate AGN host galaxies, while open symbols represent non-AGN galaxies. The broken line shows the M$_*$-M$_h$ relation defined by the overall sample. 
Right: Cumulative distribution of the offset in M$_h$ from the M$_\star$-M$_h$ relation (log M$_h$/M$_h^*$, where M$_h^*$ is the predicted M$_h$ from their M$_*$) for the MS-late, MS-early\_stellar, and MS-early\_SF galaxies.}
\label{M_Mh}
\end{figure*}

\subsection{AGN Activity}
\label{subsec:agn}
Active Galactic Nuclei (AGN) play a crucial role in galaxy evolution through their feedback effects on host galaxies \citep{Fabian12, Harrison17}. Since strong AGN emission can affect both galaxy properties and their morphological classification, it is important to examine the relationship between the presence of AGN and our galaxy subgroups.

We identify AGN hosts by requiring multiple spaxels within 0.5 Re to be classified as AGN through BPT diagnostics. This requirement helps minimize false positives that could arise from measurement uncertainties in single spaxels. 
The Figure \ref{M_SFR_AGN} shows that approximately half of the MS-early\_stellar galaxies host AGN, while AGN are rare in the MS-early\_SF galaxies. However, this distinction appears to be primarily a consequence of the different stellar mass distributions between these subgroups rather than an intrinsic characteristic of the classifications themselves.
As shown by the gray circles, the AGN fraction exhibits a strong mass dependence, with higher-mass galaxies more likely to host AGN. The stellar mass distribution of MS-early\_stellar galaxies is biased toward higher masses ($\log M_*/M_{\odot} > 10$), while MS-early\_SF galaxies tend to have a lower mass. This systematic difference in stellar mass distribution naturally accounts for the observed difference in AGN fractions between the two subgroups.

It is also important to note that our analysis excludes spaxels classified as AGN or composite to measure pure star formation activity. While this approach is necessary to avoid AGN contamination, it may lead to an underestimation of $\Sigma_\mathrm{SFR}$ (and SFR) in AGN host galaxies, particularly in their central regions. This can also affect the accuracy of our radial profile measurements. To assess the impact of this limitation, we have performed the same analyses (sSFR radial profiles and the resolved MS) using only non-AGN galaxies, as presented in Appendix A (Figure \ref{SSFRprofile_noAGN} and \ref{resolvedMS_noAGN}). We confirm that the results are basically identical to those presented in Figure \ref{SSFRprofile} and \ref{resolvedMS}, demonstrating that our main conclusions are unaffected by the AGN effects.

\subsection{Environmental Effects} 
\label{subsec:environment}

\begin{figure}
\plotone{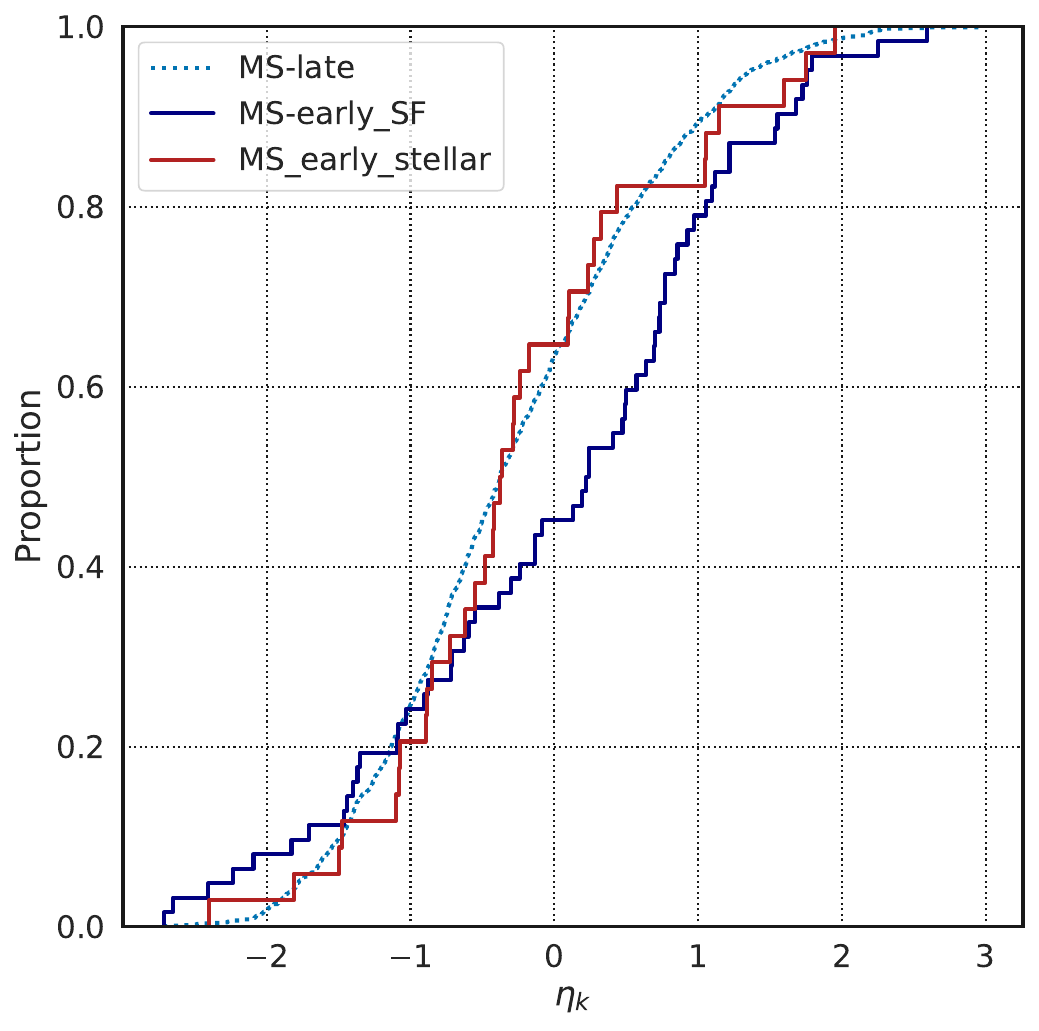}
\caption{Cumulative distribution of local density ($\eta_k$) for the three groups. The light blue dotted line represents the MS-late galaxies, while the blue and red solid lines show the MS-early\_SF and the MS-early\_stellar galaxies, respectively.}
\label{localdensity}
\end{figure}

Galaxy environment affects galaxy evolution through various physical mechanisms. High-density environments facilitate galaxy interactions and mergers that can trigger morphological transformation and enhanced star formation. In cluster environments, additional processes such as ram pressure stripping and galaxy harassment can dramatically influence galaxy properties \citep{Boselli06}.

We first examine the halo mass (M$_h$) estimates from the SDSS galaxy group catalog \citep{Tinker20, Tinker21}. This catalog identifies galaxy groups using a friends-of-friends algorithm and assigns halo masses based on the total stellar mass of the group members. 
The left panel of Figure \ref{M_Mh} shows the M$_*$--M$_h$ relation for our galaxy subgroups. The gray circles representing the full sample demonstrate a strong correlation between stellar and halo mass, consistent with previous studies \citep[e.g.,][]{Moster10, Behroozi13}.
This correlation is primarily defined by the MS galaxies, which dominate our sample. The broken line traces the peak values of the 2D kernel density estimation at each stellar mass, representing the typical M$_*$--M$_h$ relation. This line also clearly shows the underlying mass-density relation, where more massive galaxies reside in higher mass halos (i.e., denser environments). We find that the MS-early\_stellar galaxies follow this same correlation, suggesting that they inhabit environments similar to those of typical MS galaxies.

In contrast, the MS-early\_SF galaxies systematically deviate from the M$_*$--M$_h$ relation toward higher M$_h$ at fixed M$_*$. This deviation indicates that MS-early\_SF galaxies represent outliers from the typical stellar mass-environment relation, residing in environments that are denser than expected for their stellar masses. 
The right panel of Figure \ref{M_Mh} shows the cumulative distribution functions of halo mass offsets (log M$_h$/M$_h^*$) from the predicted M$_*$-M$_h$ relation, where M$_h^*$ is the predicted halo mass based on each galaxy's stellar mass. 
The cumulative distributions clearly show that MS-early\_SF galaxies exhibit much larger systematic deviations from MS-late galaxies compared to MS-early\_stellar galaxies, with MS-early\_SF galaxies showing systematic positive offsets with a median of approximately 0.6 dex. Kolmogorov-Smirnov tests comparing each MS-early subgroup with MS-late galaxies yield p-values of $2.4\times10^{-19}$ for MS-early\_SF and 0.027 for MS-early\_stellar. This also confirms that MS-early\_SF galaxies show much larger deviations than MS-early\_stellar galaxies. This systematic offset toward higher M$_h$ implies that MS-early\_SF galaxies reside in denser environments than typical MS galaxies, even after accounting for the general mass-environment correlation.

To further confirm this environmental difference, we also compare the local density distribution using the normalized local density parameter $\eta_k$ from the GEMA-VAC \citep{Argudo-Fernandez15}.
This parameter is defined as the number density within the projected distance to the 5th nearest neighbor, with higher values of $\eta_k$ indicating higher local densities. Figure \ref{localdensity} shows the $\eta_k$ distributions for each subgroup. The MS-early\_stellar galaxies exhibit a $\eta_k$ distribution similar to those of MS-late galaxies, while the MS-early\_SF galaxies show significantly higher local densities, with a median $\eta_k$ value 0.6 dex higher than that of MS-late galaxies. We perform Kolmogorov-Smirnov tests to compare the distributions of MS-early subgroups with that of MS-late galaxies. The resulting p-values are 0.905 for MS-early\_stellar and 0.002 for MS-early\_SF. At a significance level of 1 \%, the MS-early\_SF distribution differs significantly from that of MS-late galaxies.

\subsection{Formation Pathways of MS-early Galaxies} \label{subsec:formation}

Our analysis reveals that early-type galaxies on the MS consist of two distinct populations with different $\Sigma_\mathrm{SFR}$, $\Sigma_*$ geometries and environments, suggesting different formation pathways for each population.

The MS-early\_stellar galaxies maintain disk-like star formation patterns similar to MS-late galaxies while exhibiting more concentrated stellar mass distributions (Section \ref{subsec:radial_profiles}). Their environmental properties are also remarkably similar to those of MS-late galaxies, showing no evidence for environmental effects (Section \ref{subsec:environment}). These observational features suggest that MS-early\_stellar galaxies evolve primarily through internal secular processes \citep{Kormendy04, Sachdeva17}.
They appear to be simply the most bulge-dominated subset of the star-forming galaxy population, where secular evolution has gradually built up their central stellar mass while maintaining their disk-like star formation patterns. MS-early\_stellar and MS-late galaxies thus likely represent different stages along a common evolutionary sequence characterized by their degree of central mass concentration.

In contrast, MS-early\_SF galaxies show clear evidence of environmentally-driven evolution. These galaxies have centrally concentrated star formation (Section \ref{subsec:radial_profiles}) and are found in moderately denser environments than typical MS galaxies (Section \ref{subsec:environment}). Our analysis reveals that MS-early\_SF galaxies typically reside in environments with 0.6 dex higher local densities ($\eta_k$) compared to MS-late or MS-early\_stellar galaxies. This density enhancement corresponds to pair or group-scale environments rather than massive clusters or isolated field regions.

Considering the fact that MS-early\_SF galaxies are also a subsample of the MS galaxies, it is expected that they are not strong starbursts. Nevertheless, our result demonstrates that such galaxies characterized by the unique internal SF geometry are preferentially triggered by group-scale environment, suggesting that those intermediate-density environments are capable of driving the gas toward galaxy centers, triggering localized starburst-like activity without disrupting the entire galaxy. Previous studies have shown similar effects where interactions can redistribute gas within galaxies \citep[e.g.,][]{Moreno15, Blumenthal18}.
The concentrated star formation in these galaxies suggests an ongoing buildup of central stellar mass, which may eventually lead to more prominent bulge-like structures. This process appears to represent an early phase of environmentally-influenced morphological evolution that connects with the well-known trend of increasing bulge fraction in group and cluster environments \citep[e.g.,][]{Dressler80}. Additionally, the centrally concentrated star formation activity observed in MS-early\_SF galaxies may be relevant to the formation of positive age gradients found in some early-type galaxies \citep[e.g.,][]{Rawle10, Koleva11, SanRoman18, Parikh21}, where central regions contain relatively younger stellar populations than the outer regions. While we cannot directly trace the full evolutionary sequence with our data, the MS-early\_SF galaxies provide an important snapshot of environmental processes shaping galaxy structure before a stellar bulge is fully established.

In this study, we have revealed distinct subgroups \textit{amongst} MS galaxies that share similar global properties but follow different evolutionary pathways, by incorporating spatially-resolved properties. We emphasize the importance to combine the global measurements and the internal structure analysis to understand the true diversity of star-forming galaxies. While global measurements of galaxy properties characterized by e.g. the location on the MS diagram or the visual morphology are of course helpful to describe the nature of galaxies in a statistical way, the spatially integrated measurements alone cannot fully capture the diversity within the MS population. 

\section{Summary} \label{sec:summary}
We investigated the diversity within the MS galaxy population by using spatially-resolved data from the MaNGA final data release. Following the morphological classifications from the MaNGA Morphology Deep Learning catalog, we identified a sample of 1587 ``MS-late" and 97 ``MS-early" galaxies that show similar global star formation properties but different morphologies. Our main findings are as follows:

\begin{enumerate}
\item Through S\'{e}rsic profile fitting of $\Sigma_\mathrm{SFR}$ and $\Sigma_*$ maps, we found that the MS-early galaxies can be further classified into two subgroups based on their structural properties---i.e. ``MS-early\_SF" (63 galaxies) and ``MS-early\_stellar" (34 galaxies). The "MS-early\_SF" galaxies show higher S\'{e}rsic indices in $\Sigma_\mathrm{SFR}$ than $\Sigma_*$, indicating centrally concentrated star formation, while the MS-early\_stellar galaxies exhibit lower S\'{e}rsic indices in $\Sigma_\mathrm{SFR}$ than $\Sigma_*$, suggesting extended star formation similar to the MS-late galaxies.

\item We found a clear difference in the mean sSFR radial profile between the two subgroups. The MS-early\_SF galaxies show significantly enhanced central star formation with rapidly declining sSFR toward the outer regions. In contrast, the MS-early\_stellar galaxies exhibit profiles similar to the MS-late galaxies, characterized by suppressed central star formation and relatively flat sSFR distributions in their outer regions.

\item We studied the resolved SFR--$M_*$ relation, and found that the MS-early\_SF galaxies show steeper slopes in their $\Sigma_*$-$\Sigma_\mathrm{SFR}$ relations compared to the MS-late or MS-early\_stellar galaxies, demonstrating their intense central star formation activity. On the other hand, the slope of the resolved main sequence for the MS-early\_stellar galaxies is similar to that of the MS-late galaxies, although the MS-early\_stellar galaxies show more concentrated stellar mass distribution than MS-late galaxies.

\item Approximately half of the MS-early\_stellar galaxies host AGNs, while AGNs are rare in the MS-early\_SF galaxies. We note that this difference would reflect their different stellar mass distributions (the MS-early\_stellar galaxies tend to be more massive than MS-early\_SF). However, our conclusions are unchanged even if we use only non-AGN galaxies.

\item The MS-early\_SF galaxies tend to reside in different environment than MS-early\_stellar or MS-late galaxies. The MS-early\_stellar galaxies have similar halo masses and local densities to those of the MS-late galaxies. In contrast, the MS-early\_SF galaxies show systematic environmental differences. They typically have 0.6 dex higher halo masses at fixed stellar mass (based on median values) and reside in environments with 0.6 dex higher local densities compared to MS-late galaxies. These significant environmental differences suggest distinct formation mechanisms between these populations.

\item Our results imply that there exist two different evolutionary pathways among the MS galaxies with early-type morphologies. The MS-early\_stellar galaxies share similar environments to MS-late galaxies and maintain disk-like star formation patterns, suggesting they evolve through internal processes that gradually build up central mass. The MS-early\_SF galaxies, in contrast, reside in denser group-scale environments and show centrally concentrated star formation. This environmental connection suggests that interactions in these denser regions drive gas toward galaxy centers, triggering intense localized star formation. While these galaxies lack stellar bulges, their concentrated star formation activity may represent an early phase of environmentally-driven morphological evolution, potentially connecting to the higher bulge fractions observed in group and cluster environments.
\end{enumerate}

We emphasize again that, while the positions on the SFR--$M_*$ diagram can provide a basic information to characterize the nature of galaxies, there exist diverse galaxy populations even at a fixed stellar mass and SFR. By combining the global and spatially-resolved properties of galaxies, we now identified distinct subgroups among the MS galaxies---galaxies sharing similar global star formation properties but likely followed via different evolutionary pathways. Our results demonstrate that spatially-resolved information from IFS observations is critical for understanding the full picture of galaxy evolution.

\begin{acknowledgements}
We thank the anonymous referee for careful reading
and useful comments that helped to improve our paper.
Funding for the Sloan Digital Sky 
Survey IV has been provided by the 
Alfred P. Sloan Foundation, the U.S. 
Department of Energy Office of 
Science, and the Participating 
Institutions. 
SDSS-IV acknowledges support and 
resources from the Center for High 
Performance Computing  at the 
University of Utah. The SDSS 
website is www.sdss4.org.
SDSS-IV is managed by the 
Astrophysical Research Consortium 
for the Participating Institutions 
of the SDSS Collaboration including 
the Brazilian Participation Group, 
the Carnegie Institution for Science, 
Carnegie Mellon University, Center for 
Astrophysics | Harvard \& 
Smithsonian, the Chilean Participation 
Group, the French Participation Group, 
Instituto de Astrof\'isica de 
Canarias, The Johns Hopkins 
University, Kavli Institute for the 
Physics and Mathematics of the 
Universe (IPMU) / University of 
Tokyo, the Korean Participation Group, 
Lawrence Berkeley National Laboratory, 
Leibniz Institut f\"ur Astrophysik 
Potsdam (AIP),  Max-Planck-Institut 
f\"ur Astronomie (MPIA Heidelberg), 
Max-Planck-Institut f\"ur 
Astrophysik (MPA Garching), 
Max-Planck-Institut f\"ur 
Extraterrestrische Physik (MPE), 
National Astronomical Observatories of 
China, New Mexico State University, 
New York University, University of 
Notre Dame, Observat\'ario 
Nacional / MCTI, The Ohio State 
University, Pennsylvania State 
University, Shanghai 
Astronomical Observatory, United 
Kingdom Participation Group, 
Universidad Nacional Aut\'onoma 
de M\'exico, University of Arizona, 
University of Colorado Boulder, 
University of Oxford, University of 
Portsmouth, University of Utah, 
University of Virginia, University 
of Washington, University of 
Wisconsin, Vanderbilt University, 
and Yale University.
YK acknowledge support from JSPS KAKENHI grant No.23H01219 and JSPS Core-to-Core Program (grant No.JPJSCCA20210003).
\software{Astropy \citep{astropy:2013, astropy:2018, astropy:2022},
NumPy \citep{2020NumPy-Array},
SciPy \citep{2020SciPy-NMeth},
Pandas \citep{reback2020pandas},
Matplotlib \citep{Hunter07},
JAX \citep{jax2018github},
NumPyro \citep{phan2019composable, bingham2019pyro},
Marvin \citep{Cherinka19},
}
\end{acknowledgements}


\bibliographystyle{aasjournal}

\appendix
\restartappendixnumbering
\section{Analysis excluding AGN host galaxies} \label{appendix:noAGN}
\begin{figure*}
\plotone{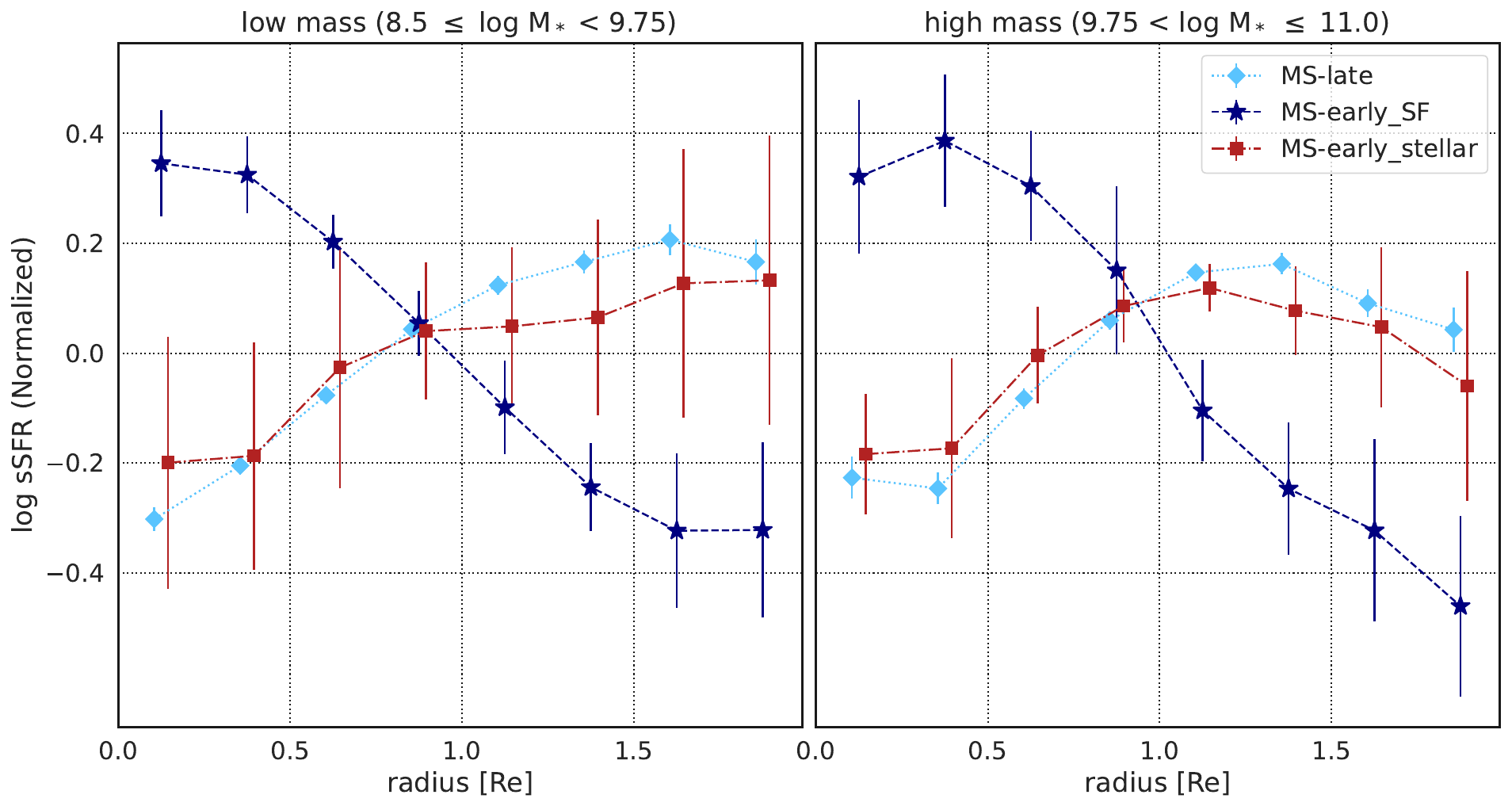}
\caption{
The mean sSFR radial profiles for non-AGN galaxies in the three groups: MS-late (light blue diamonds), MS-early\_SF (navy stars), and MS-early\_stellar (red squares). As in Figure \ref{SSFRprofile}, the error bars represent the 95\% confidence intervals of the mean, assuming a t-distribution, and profiles are shown separately for low-mass ($8.5 \leq \log(M/M_\odot) < 9.75$; left panel) and high-mass ($9.75 \leq \log(M/M_\odot) < 11.0$; right panel) galaxies.
}
\label{SSFRprofile_noAGN}

\plotone{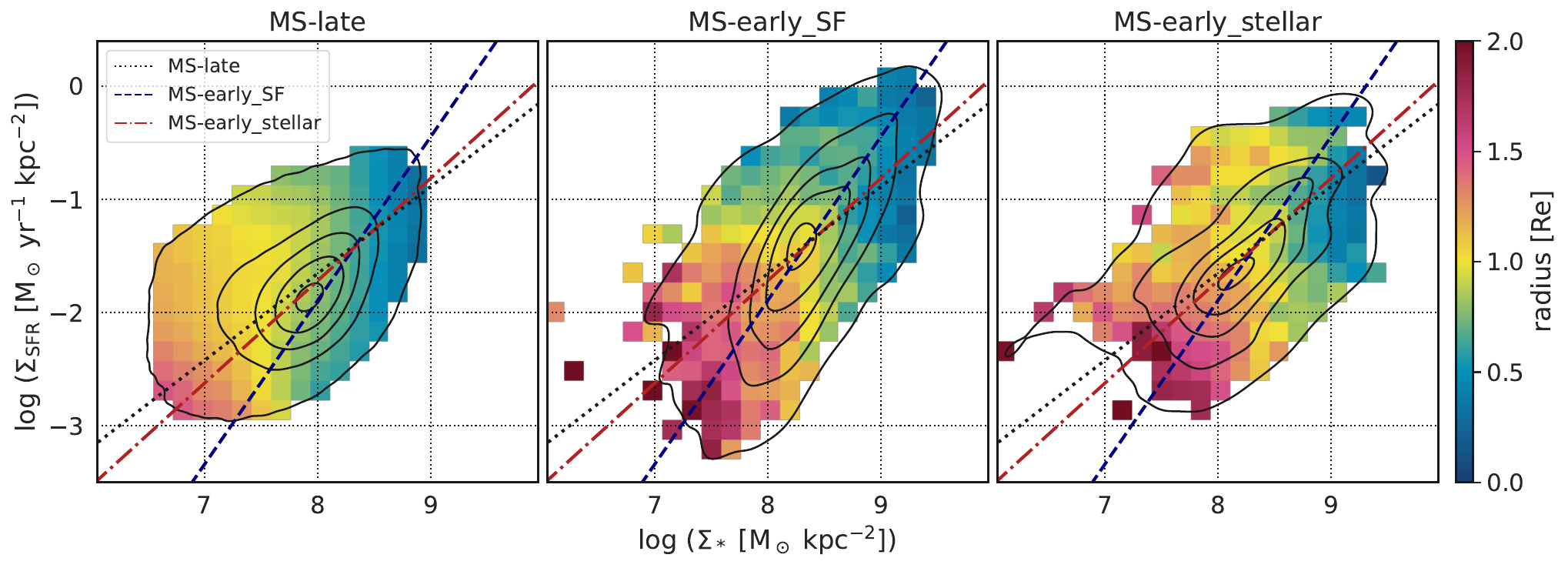}
\caption{
The resolved main sequence for non-AGN galaxies in the three groups. From left to right, the panels show the distributions for MS-late, MS-early\_SF, and MS-early\_stellar galaxies. As in Figure \ref{resolvedMS}, contours represent the 2D histogram of spaxel values, and the overlaid colormaps correspond to the mean radius [R$_e$] at each position. The three lines indicate the results of orthogonal fitting to the distributions. The black dotted line corresponds to the MS-late, the blue dashed line corresponds to the MS-early\_SF, and the red dot-dashed line corresponds to the MS-early\_stellar.
}
\label{resolvedMS_noAGN}
\end{figure*}

To verify that our results are not significantly affected by AGN in our analysis, we reproduce our main result's figures using only non-AGN galaxies. Figure \ref{SSFRprofile_noAGN} shows the mean sSFR radial profiles, and Figure \ref{resolvedMS_noAGN} presents the resolved main sequence relations for galaxies without AGN. The patterns observed in these figures are nearly identical to those presented in Figures \ref{SSFRprofile} and \ref{resolvedMS}, respectively, confirming that our main conclusions are robust against potential AGN effects.

\end{document}